\documentclass[aps,showpacs,11pt]{revtex4}
\usepackage{dcolumn}
\usepackage{graphicx}
\usepackage{amsmath}
\usepackage{amsfonts}
\usepackage{amssymb}
\usepackage{psfrag}
\usepackage{wrapfig}
\usepackage{subfigure}
\usepackage{makeidx}
\usepackage{bm}
\usepackage{epsf}
\usepackage{hyperref}
\usepackage{color}
\usepackage{float}

\begin{document}

\title{Superradiant instability of extremal brane-world Reissner-Nordstr\"{o}m black holes to charged scalar perturbations}

\author{Shao-Jun Zhang, Bin Wang}
\affiliation{CAA and Department of Physics and Astronomy, Shanghai Jiao Tong University, Shanghai 200240, China}
\author{Elcio Abdalla}
\affiliation{Instituto de F$\acute{i}$sica, Universidade de S$\tilde{a}$o Paulo, C.P.66.318, CEP 05315-970, S$\tilde{a}$o Paulo, Brazil}

%\email{$^a$sjzhang84@sjtu.edu.cn}\\
%\email{$^b$Wang$_$b@sjtu.edu.cn}\\
%\email{$^c$eabdalla@fma.if.usp.br}

\date{\today}

\begin{abstract}
\indent

We examine the stability of extremal braneworld  Reissner-Nordstr\"{o}m black holes under massive charged scalar
perturbations. We show that similar to the four-dimensional case, the superradiant amplification can occur in
braneworld charged holes.  More interestingly we find that when the spacetime dimension is higher than four, a
trapping potential well can emerge outside the braneworld black hole. This potential well is the extra dimensional
effect and does not exist in four dimensions. It triggers the superradiant instability  of the extremal braneworld
charged  holes.

\end{abstract}



\maketitle

Starting from the influential study by Regge and Wheeler \cite{Regge:1957a}, the stability of black holes has been
investigated over half a century. It has been demonstrated that most black holes are stable under  various types of
perturbations (for a recent review see for example \cite{Konoplya:2011a}), which shows that the black hole is
realizable in practice and is not just a mathematical curiosity.

In Einstein gravity, the Kerr family exhausts the black hole solutions of the Einstein equations in the vacuum.
The Kerr black hole is rotating, which is a realistic model to describe astrophysics.  The Kerr solution was
shown to be stable under dynamical perturbations
\cite{Press:1973a,Teukolsky:1972a,Teukolsky:1973a,Hod:1998a,Hod:1999a,Hod:1999b,Barack:1999a,Hod:1999c,Gleiser:2007a,Tiglio:2007a,Hod:2008a,Hod:2008b,Zenginoglu:2009a,Amsel:2009a,Cardoso:2004a,Cardoso:2004b}.
However, such a conclusion will be changed due to the superradiance effect \cite{Press:1972a}.  Considering a wave of
the form $e^{-i\omega t+i k\phi}$ incident upon a Kerr black hole with the angular velocity $\Omega$ , if the frequency
of the incident wave satisfies $\omega<k\Omega$,  the scattered wave will be amplified. This means that the energy
radiated away to infinity can exceed the energy of the initial perturbation by extracting rotational energy of the
black hole. This superradiance can affect the classical fields and influence the stability of the background
configuration.  If the Kerr black hole is enclosed inside a spherical mirror, the initial perturbation can get
successfully amplified near the black hole event horizon and reflected back at the mirror, thus creating an instability.
This is the black hole bomb devised in  \cite{Press:1972a}.  Besides the artificial mirror, one can devise a natural
wall if one considers massive fields
\cite{Cardoso:2004a,Cardoso:2004b,Damour:1976a,Zouros:1979a,Detweiler:1980a,Furuhashi:2004a,Dolan:2007a,Cardoso:2005a,Hod:2009a,Beyer:2011a,Hod:2012a}.
For a massive field, the field amplified by superradiance gets trapped in the effective potential well, thus triggering
the instability.   The bound state due to the potential well plays the same role as the artificial mirror, what results
in the instability.

The  Reissner-Nordstr\"{o}m (RN) black holes share many common features with the Kerr black holes. The stability of the
RN black hole  under neutral perturbations was disclosed long time ago~\cite{Moncrief:1974a,Moncrief:1974b}.
The analogous superradiant phenomena was also observed in the RN black hole \cite{Bekenstein:1973a}. Considering the
charged perturbation of a bosonic field, if the frequency of the incident wave $\omega$ satisfies the relation
\begin{eqnarray}
\omega<e \Phi,\label{4d-SRregime}
\end{eqnarray}
where $e$ is the charge of the incident field and $\Phi$ is the electric potential of the RN black hole, then
superradiance occurs by extracting the Coulomb energy and the electric charge from the RN black hole. Can this
superradiance trigger the instability as disclosed in the Kerr black hole?  Recently, this problem was examined
in \cite{Hod:2013a,Hod:2013b}.  It was found that for massive charged scalar perturbations, there does not exist a
trapping potential well separated from the black hole horizon by a potential barrier outside the black hole. This shows
that for the RN black hole there is no bound state to mimic the artificial mirror as in the Kerr black hole case to trigger
the superradiant instability.

The stability analysis has been performed mainly in the four-dimensional black hole backgrounds. Recently, extensive studies of d-dimensional cases led to the fact that black branes and strings are generally unstable against a certain sector of gravitational perturbations \cite{Harmark:2007a}.  More interestingly, it was observed that the usual stability result does not hold in the RN de
Sitter black holes for spacetime dimensions larger than six \cite{Konoplya:2008a,Cardoso:2010a}.  It is of great interest to
examine how is the extra dimensional effect on the superradiant instability. This is the main purpose of the present work. We extend
the discussion in usual four-dimensional RN black holes \cite{Hod:2013a,Hod:2013b} to the brane-world black holes \cite{Arkani-Hamed:1998a,Antoniadis:1998a,Randall:1999a,Randall:1999b}. Notice that these are black holes in the brane, that is, we do not consider perturbations that run into the extra dimensions. The $(4+n)$-dimensional RN black hole was described in  \cite{Myers:1986a} in the form
\begin{eqnarray}
ds^2&=&-f(r) dt^2 + \frac{1}{f(r)} dr^2 +r^2 d\Omega_{n+2}^2,\nonumber\\
f(r)&=&1-\frac{2\mu}{r^{n+1}}+\frac{q^2}{r^{2(n+1)}},\label{RN}
\end{eqnarray}
where  $n$ denotes the number of compact extra dimensions, the parameters $\mu$, $q$ are related to the mass $M$ and charge $Q$ of the black hole through the relations
\begin{eqnarray}
\mu&=&\frac{8\pi G M}{(n+2) V_{n+2}},\nonumber\\
q^2&=&\frac{8\pi G Q^2}{(n+2)(n+1)},\label{mass-charge}
\end{eqnarray}
and the volume of the $(n+2)$-dimensional sphere is  $V_{n+2}=\frac{2\pi^{(n+3)/2}}{\Gamma\left(\frac{n+3}{2}\right)}$. $d\Omega_{n+2}^2$ describes the corresponding line-elements of the $(n+2)$-dimensional unit sphere,
\begin{eqnarray}
d\Omega_{n+2}^2=d\theta_{n+1}^2 +\sin^2\theta_{n+1} \left(d\theta_n^2 +\sin^2\theta_n\left(\ldots +\sin^2\theta_2\left(d\theta_1^2+\sin^2\theta_1 d\phi^2\right)\ldots\right)\right).\label{unit-sphere}
\end{eqnarray}
%\begin{eqnarray}
%d\Omega_{n+2}^2= d\theta^2 +\sin^2\theta d\phi^2 +\sin^2\theta \sin^2\phi\left( d\theta_1^2 +\sin^2\theta_1 d\theta^2 +\cdots +\sin^2\theta_1 \cdots \sin^2\theta_{n-1} d\theta_n^2\right).
%\end{eqnarray}
In the above, $\phi \in [0,2\pi]$ and $\theta_i \in [0,\pi] \ (i=1,\ldots,n+1)$. We use $n$ additional azimuthal coodinates $\theta_i
\ (i=2,\ldots,n+1)$ to denote the $n$ compact extra dimensions.
The Maxwell field is
\begin{eqnarray}
A=h(r) dt,\hspace{1cm}
h(r)=-\sqrt{\frac{1}{8\pi G} \frac{n+2}{n+1}} \frac{q}{r^{n+1}}.\label{electric}
\end{eqnarray}

We concentrate on the charged perturbation of a bosonic field.  Considering that the standard model fields live on the brane while
only gravity can travel into the extra dimensions in the braneworld scenario, we focus on the braneworld metric induced
by a $(4 + n)$-dimensional RN extreme spacetime. This simply follows by fixing the values of the extra angular azimuthal
coordinates, i.e. $\theta_i =\frac{\pi}{2}$, for $i=2,\ldots, n+1$, which leads to the projection of the $(4+n)$-dimensional RN metric  (\ref{RN}) on a four-dimensional slice playing the role of our four-dimensional world. The induced metric then has the form
\begin{eqnarray}
ds^2=-f(r) dt^2 + \frac{1}{f(r)} dr^2 +r^2 (d\theta^2+\sin^2\theta d\phi^2)\quad ,\label{induceRN}
\end{eqnarray}
where the subscript ``1" from the remaining azimuthal coordinate has been dropped. Note that the metric function $f(r)$ remains unchanged during the projection and is still given by (\ref{RN}). Since the Maxwell field (\ref{electric}) totally lives on the brane, it remains the same under the projection.
We show that with the extra dimensional influence, the effective potential out of the black hole can have
a potential well to act as the natural mirror to trap the superradiant wave and thus trigger the instability.

In this paper, we only consider the extremal case, satisfying the condition $\mu=q$. Thus the metric coefficient function $f(r)$ is
given by the expression
\begin{eqnarray}
f(r) = \left(1-\frac{\mu}{r^{n+1}}\right)^2\quad ,\label{extremeRN}
\end{eqnarray}
and the degenerate horizon is located at $r_H=\mu^{1/(n+1)}$.

We consider a massive charged scalar living on the brane. Its dynamics in the braneworld RN extreme black hole is
governed by the Klein-Gordon equation
\begin{eqnarray}
\left[(\nabla_\mu-i e A_\mu) (\nabla^\mu-i e A^\mu)-m^2\right]\Psi=0,\label{EoM}
\end{eqnarray}
where $e$ and $m$ are the charge and mass of the field, respectively. We do the decomposition of the field as
\begin{eqnarray}
\Psi_{lk} (t,r,\theta,\phi)=e^{-i\omega t} R_{lk}(r) S_{lk} (\theta) e^{-i k\phi},\label{decomposition}
\end{eqnarray}
where $\omega$ is the conserved energy of the mode, $l$ is spherical harmonic index, and $k$ is the azimuthal harmonic
index with $-l\leq k\leq l$. In the following, for brevity, we will omit the indices $l$ and $k$. With the decomposition,
we can obtain the radial Klein-Gordon equation
\begin{eqnarray}
r^2 f(r) \frac{d}{dr}\left(r^2 f(r) \frac{dR}{dr}\right)+UR=0,\label{radial}
\end{eqnarray}
with
\begin{eqnarray}
U=\left(\omega+e h(r)\right)^2 r^4-f(r) \left[m^2 r^4+l(l+1)r^2\right].\label{U}
\end{eqnarray}

It is convenient to define a new radial function $\psi$ by
\begin{eqnarray}
\psi\equiv r f(r)^{1/2} R.\label{psi}
\end{eqnarray}
Then the radial equation (\ref{radial}) can be written in the form of a Schrodinger-like wave equation,
\begin{eqnarray}
\frac{d^2\psi}{dr^2} + \left(\omega^2-V_{eff}\right)\psi=0,\label{radial1}
\end{eqnarray}
where the effective potential is
\begin{eqnarray}
V_{eff}=\omega^2-\frac{1}{f(r)^2} \left[\frac{U}{r^4}+\frac{1}{4}  \left(\frac{df(r)}{dr}\right)^2-\frac{f(r)}{r}
\frac{df(r)}{dr}-\frac{f(r)}{2} \frac{d^2f(r)}{dr}\right].\label{potential}
\end{eqnarray}
In the following, we set $8\pi G=c=\hbar=1$. Substituting the expressions of $f(r)$ and $h(r)$ from  (\ref{extremeRN}) and
(\ref{electric}) into the above equation, we get
\begin{eqnarray}
V_{eff}=\omega^2&+&\frac{1}{\left(r^{n+1}-\mu\right)^4}\bigg[(m^2-\omega^2) r^{4n+4}+l(l+1)r^{4n+2}+2\left(\sqrt{
\frac{n+2}{n+1}}e\omega-m^2\right)\mu r^{3n+3}\nonumber\\
&&\qquad\qquad\qquad-\left[n(n+1)+2l(l+1)\right]\mu r^{3n+1}+\left(m^2-\frac{n+2}{n+1}e^2\right)\mu^2 r^{2n+2}\nonumber\\
&&\qquad\qquad\qquad+\left[3n(n+1)+l(l+1)\right]\mu^2 r^{2n}
-3n(n+1)\mu^3 r^{n-1}\nonumber\\
&&\qquad\qquad\qquad+n(n+1) \mu^4 r^{-2}\bigg].\label{potential1}
\end{eqnarray}
As $r\rightarrow r_H$, we have $V_{eff}\rightarrow -\frac{r_H^4 (\omega-e\Phi)^2}{(n+1)^4 (r-r_H)^4}$, where $\Phi=-h(r_H)$
is the electric potential at the horizon.  When $r \rightarrow \infty$, $V_{eff}\rightarrow m^2$.

 In a scattering experiment,  (\ref{radial1}) has the following asymptotic behavior with $\omega^2>m^2$ 
\begin{eqnarray}
\psi \sim \left\{
\begin{array}{ll}
T e^{-i\sigma (r-r_H)^{-1}} & \quad{\rm as} \  r\rightarrow r_H\quad ,\\
R e^{i\sqrt{\omega^2-m^2}r}+e^{-i\sqrt{\omega^2-m^2}r} & \quad {\rm as}\  r\rightarrow \infty\quad ,
\end{array}
\right. \label{bny}
\end{eqnarray}
 The boundary conditions correspond to an incident wave of unit amplitude, $e^{-i\sqrt{\omega^2-m^2}r}$, coming from $+\infty$ and giving rise to a reflected wave of
amplitude $R$ going back to $+\infty$ and a transmitted wave of amplitude $T$ at the horizon.  We also define the constant
$\sigma \equiv \frac{(e\Phi-
\omega) r_H^2}{(n+1)^2}$. At the black hole horizon, the boundary behavior of the radial equation (\ref{bny}) is that of a
purely ingoing wave.

Considering that the effective potential is real, the complex conjugate of the solution $\psi$ satisfying the boundary
conditions~(\ref{bny}) will satisfy the complex-conjugate boundary conditions:
\begin{eqnarray}
\psi^\ast \sim \left\{
\begin{array}{ll}
T^\ast e^{i\sigma (r-r_H)^{-1}} & \quad{\rm as} \  r\rightarrow r_H,\\
R^\ast e^{-i\sqrt{\omega^2-m^2}r}+e^{i\sqrt{\omega^2-m^2}r} & \quad {\rm as}\  r\rightarrow \infty.
\end{array}
\right. \label{bny1}
\end{eqnarray}

Because the two solutions $\psi$ and $\psi^\ast$ are linearly independent, then their Wronskian, $W(\psi,\psi^\ast)\equiv
\psi \frac{d}{dr}\psi^\ast-\psi^\ast \frac{d}{dr}\psi$, is a constant independent of $r$. Evaluating the Wronskian at the
horizon and infinity respectively, we get
\begin{eqnarray}
W(r\rightarrow r_H)&=&-\frac{2i\sigma}{(r-r_H)^2} |T|^2,\nonumber\\
W(r\rightarrow \infty)&=&-2i\sqrt{\omega^2-m^2} (|R|^2-1). \label{Wronskian}
\end{eqnarray}
By equating the two values, we get
\begin{eqnarray}
|R|^2 =1+\frac{\sigma}{(r-r_H)^2 \sqrt{\omega^2-m^2}}|T|^2. \label{reflect}
\end{eqnarray}
Now, we can see that if $\sigma>0$, we have $|R|^2>1$. This means that one gets back more than one threw in, and superradiant
phenomena occurs. So we get the condition to occur the superradiance,
\begin{eqnarray}
\omega<e\Phi, \label{SRregime}
\end{eqnarray}
which has the same form as in the four-dimensional case, and does not change in the braneworld picture. However, the
dimensional influence hides in the electric potential.

To see whether the superradiance will cause the instability of the braneworld black hole spacetime, we need to check
whether there exists a potential well outside the horizon to trap the reflected wave. If the potential well exists,
the superradiant instability will occur  and the wave will grow exponentially over time near the black hole to make
the background braneworld black hole unstable.

Now, we analyze the behavior of the effective potential.  From Eq.~(\ref{potential}), we can get the derivative of the
effective potential as
\begin{eqnarray}
V_{eff}'(r;\mu,m,e,\omega,l,n)=&&\frac{1}{\left(r^{n+1}-\mu\right)^5}\bigg[-2l(l+1)r^{5n+2}
-2(n+1)\left[m^2+\left(e \sqrt{\frac{n+2}{n+1}}-2\omega\right)\omega\right]\mu r^{4n+3}\nonumber\\
&&+\left[n(n+1)(n+3)-2(n-2)l(l+1)\right]\mu r^{4n+1}\nonumber\\
&&+\left[2(2n-1)l(l+1)-n(n+1)(3n+11)\right]\mu^2 r^{3n}\nonumber\\
&&+2\left[2(n+1)m^2 + (n+2)e^2-3\sqrt{(n+1)(n+2)}e\omega\right]\mu^2 r^{3n+2}\nonumber\\
&&+2\left[(n+2)e^2-(n+1)m^2\right]\mu^3 r^{2n+1}+n\left[3(n+1)(n+5)-2l(l+1)\right]\mu^3r^{2n-1}\nonumber\\
&&-n(n+1)(n+9)\mu^4 r^{n-2}+2n(n+1)\mu^5 r^{-3}\bigg]\quad .\label{potentialprime}
\end{eqnarray}
We denote the roots of $V_{eff}' (r)=0$ which are larger than $r_H$ by $\{r_1, r_2, \cdots, r_N\}$ with $r_H<r_1\leq r_2
\leq \cdots \leq r_N.$ As $r\rightarrow r_H$, we have
\begin{eqnarray}
V_{eff}' (r\rightarrow r_H) = \frac{4 (n+1)\mu^{(5n+4)/(n+1)}}{\left(r^{n+1}-\mu\right)^5} \left(\omega-e\Phi\right)^2.\label{nearhorizon}
\end{eqnarray}
As long as $\omega\not = e\Phi$, which includes the superradiant regime (\ref{SRregime}) we considered, the derivative of the potential near the horizon is always positive, which means that
the first zero point $r_1$ of $V_{eff}' (r)$ corresponds to the maximum of the effective potential $V_{eff}$. This can
be easily checked by calculating the sign of $V_{eff}'' (r)$. In \cite{Hod:2013b}, it was shown analytically that for the $n=0$
case there only exists one root of $V_{eff}' (r)=0$ which is larger than $r_H$. Thus in four dimensions, there is only a
potential barrier (and no potential well) outside the black hole so that there is no superradiant instability. However,
when $n\geq 1$, the situation becomes very different. From (\ref{potentialprime}), we can see that when $n\geq 1$, the equation
becomes a rather high order equation. It is hard to analyze its roots analytically. We will count on numerical analysis.

In table I, we list the numerical roots of $V_{eff}' (r)=0$  and the corresponding $V_{eff}$ for $1\leq n\leq 6$. We fix
the parameters $\mu=1$ and $l=0$, so that the horizon locates at $r_H=1$.  We are interested in solutions of the radial equation (\ref{radial1}) with the physical boundary conditions of purely ingoing waves at the horizon and a decaying solution at spatial infinity.  A bound state decaying exponentially at spatial infinity is characterized by $\omega^2<m^2$.  We choose the parameters ($e\Phi, m$
and $\omega$) to meet this condition together with the super-radiant condition (\ref{SRregime}).  We observe  that there exists a negative minima of the potential well. If we choose other parameters,
we can also obtain a positive minima of the potential well.  From the table, we can see that there may exist
more than one root for $n\geq 1$. The first root $r_1$ corresponds to the location of a potential barrier, the
second root $r_2$ corresponds to the position of a potential well, and the third one $r_3$ is the place of a potential
barrier again. This can be confirmed by checking the sign of $V_{eff}''$ at these points. These results imply that, when
$n\geq 1$, there may exist a potential well which is separated from the horizon by a potential barrier. Then the two
conditions to trigger superradiant instability can be satisfied simultaneously. This is very different from that in
the four-dimensional case~\cite{Hod:2013b}.

\begin{table}[H]
\begin{center}
\begin{tabular}{|c|c|c|c|c|c|c|}
\hline
& $n=1$ & $n=2$ & $n=3$ & $n=4$ & $n=5$ & $n=6$\\
\hline
$e\Phi$ & $0.88$ & $1$ & $1$ & $1$ & $1$ & $1$ \\
\hline
$m$ & $0.86$ & $0.9$ & $1.2$ & $1.8$ & $2$ & $2$\\
\hline
$\omega$ & $0.85$ & $0.85$ & $0.85$ & $0.85$ & $0.85$ & $0.85$\\
\hline
$r_1$ & $1.01176$ & $1.04196$ & $1.02797$ & $1.01674$ & $1.01327$ & $1.01214$\\
\hline
$r_2$ & $1.16282$ & $1.12098$ & $1.07759$ & $1.08363$ & $1.055$ & $1.03202$ \\
\hline
$r_3$ & $6.59754$ & $4.75378$ & $3.23743$ & $2.27369$ & $2.41529$ & $2.8146$\\
\hline
$V_{eff}(r_1)$ & $975.3594$ & $12.4182$ & $17.916$ & $164.914$ & $173.854$ & $34.6291$\\
\hline
$V_{eff}(r_2)$ & $-1.19747$ & $-4.10825$ & $-10.0049$ & $-6.85324$ & $-20.1684$ & $-58.4737$\\
\hline
$V_{eff}(r_3)$ & $0.740589$ & $0.811597$ & $1.44511$ & $3.26462$ & $4.00854$ & $4.00108$\\
\hline
\end{tabular}
\end{center}
\caption{Roots of $V_{eff}' (r) =0$ and the corresponding $V_{eff}$ for $1\leq n\leq 6$.}
\end{table}

In table II, we fix the parameters: $\mu=1, l=0, e\Phi=1, m=2$ and $\omega=0.85$, and list the roots and the corresponding
effective potential for $1\leq n\leq 6$.  From the table, we can see that as $n$ increases, a potential well appears and
becomes deeper and deeper, and the first and second potential barrier becomes lower and lower.

\begin{table}[H]
\begin{center}
\begin{tabular}{|c|c|c|c|c|c|c|c|}
\hline
& n=0 & $n=1$ & $n=2$ & $n=3$ & $n=4$ & $n=5$ & $n=6$ \\
\hline
$r_1$ & $1.07367$ & $1.03639$ & $1.02441$ & $1.01861$ & $1.01529$ & $1.01327$ & $1.01214$\\
\hline
$r_2$ & $-$ & $-$ & $-$ & $-$ & $1.10994$ & $1.055$ & $1.03202$\\
\hline
$r_3$ & $-$ & $-$ & $-$ & $-$ & $1.98135$ & $2.41529$ & $2.8146$\\
\hline
$V_{eff}(r_1)$ & $551.515$ &$526.309$ & $474.566$ & $397.286$ & $296.166$ & $173.854$ & $34.6291$\\
\hline
$V_{eff}(r_2)$ & $-$ & $-$ & $-$ & $-$ & $-1.1807$ & $-20.1684$ & $-58.4737$\\
\hline
$V_{eff}(r_3)$ & $-$ & $-$ & $-$ & $-$ & 4.0624 & 4.00854 & 4.00108\\
\hline
\end{tabular}
\end{center}
\caption{Roots of $V_{eff}' (r) =0$ and the corresponding $V_{eff}$ for $1\leq n\leq 6$ with fixed parameters: $e\Phi=1, m=2, \omega=0.85$. Here "$-$" means not exist.}
\end{table}

In conclusion, we have investigated the possible existence of superradiant instability for extreme charged RN brane-world
black hole due to the charged massive perturbations on the brane. We have shown that in contrast with the four-dimensional
charged black hole, there exists a trapping potential well when we consider the extra dimensional contribution. Thus, the
wave can be trapped in the well and amplified by superradiance. This triggers the superradiant instability in the charged
brane-world black hole.

\section*{Acknowledgments}

This work is supported by the National Natural Science Foundation of China.

\end{document}